\documentclass[manuscript]{emulateapj}
\usepackage{subfigure}

\shorttitle{Two-step evolution of a rising flux rope resulting in a confined solar flare}
\shortauthors{Yang et al.}

\journalinfo{Accepted for publication in ApJ}
\submitted{ApJ, accepted 2019 May 2}

\begin{document}

\title{Two-step evolution of a rising flux rope resulting in a confined solar flare}

\author{Shuhong Yang\altaffilmark{1,2}, Jun Zhang\altaffilmark{1,2}, Qiao Song\altaffilmark{3}, Yi Bi\altaffilmark{4}, and Ting Li\altaffilmark{1,2}}

\altaffiltext{1}{CAS Key Laboratory of Solar Activity, National
Astronomical Observatories, Chinese Academy of Sciences, Beijing
100101, China; shuhongyang@nao.cas.cn}

\altaffiltext{2}{School of Astronomy and Space Science, University
of Chinese Academy of Sciences, Beijing 100049, China}

\altaffiltext{3}{Key Laboratory of Space Weather, National Center for Space Weather, China Meteorological Administration, Beijing 100081, China}

\altaffiltext{4}{Yunnan Observatories, Chinese Academy of Sciences, Kunming 650011, China}

\begin{abstract}

Combining the Solar Dynamics Observatory and the New Vacuum Solar Telescope observations, we study a confined flare triggered by a rising flux rope within the trailing sunspots of active region 12733. The flux rope lying above the sheared polarity inversion line can be constructed through magnetic extrapolation but could not be detected in multi-wavelength images at the pre-flare stage. The conspicuous shearing motions between the opposite-polarity fields in the photosphere are considered to be responsible for the flux rope formation. The maximum twist of the flux rope is as high as $-$1.76, and then the flux rope rises due to the kink instability. Only when the flare starts can the flux rope be observed in high-temperature wavelengths. The differential emission measure results confirm that this flux rope is a high-temperature structure. Associated with the rising flux rope, there appear many post-flare loops and a pair of flare ribbons. When the rising flux rope meets and reconnects with the large-scale overlying field lines, a set of large-scale twisted loops are formed, and two flare ribbons propagating in opposite directions appear on the outskirts of the former ribbons, indicating that the twist of the flux rope is transferred to a much larger system. These results imply that the external reconnection between the rising flux rope and the large-scale overlying loops plays an important role in the confined flare formation.

\end{abstract}


\section{Introduction}

Solar flares as one of the most energetic phenomena on the Sun have been extensively studied for many years (e.g., Priest \& Forbes 2002; Janvier et al. 2015; Tian et al. 2014; Yang et al. 2017), and magnetic reconnection is deemed to be an efficient way for the sudden energy release (Zweibel \& Yamada 2009; Yang et al. 2015; Wyper et al. 2017). Magnetic flux ropes appearing as filaments if filled with dark chromospheric material are thought to play a crucial role in the initiation of solar flares. Flux ropes could become unstable under some conditions, such as kink instability (T{\"o}r{\"o}k \& Kliem 2005), tether-cutting reconnection (Moore et al. 2001), and breakout reconnection (Antiochos et al. 1999). When a flux rope begins to rise, the overlying loops are stretched and a current sheet is created between the oppositely directed field lines beneath the rising rope (Shibata et al. 1995; Lin \& Forbes 2000). Consequently magnetic reconnection takes place and free energy is released, forming a solar flare (Masuda et al. 1994).

Solar flares associated with coronal mass ejections are called eruptive flares, and the others are confined flares (Svestka \& Cliver 1992; Schrijver 2009; Yang et al. 2014a; Yang \& Zhang 2018). As revealed by observations and simulations, the strong overlying magnetic loops forming a confining cage can prevent the outward escape of flux ropes, and thus result in confined flares (T{\"o}r{\"o}k \& Kliem 2005; Guo et al. 2010; Joshi et al. 2015; Liu et al. 2016a; Amari et al. 2018). In addition, if the tension force of the guide field is strong enough, it is also hard for a flux rope to erupt successfully (Myers et al. 2015). For a failed eruption, the rising flux rope or filament decelerates and reaches to a certain maximum height, after which the dark filament material if exists will drain back to the solar surface (e.g., Ji et al. 2003).

In this paper, we report the observations of a rising flux rope with a two-step evolution process involved in a confined flare, and propose that the external reconnection between the rising flux rope and a part of overlying field lines plays an important role in the confined flare formation.

\section{Observations and Data Analysis}

\begin{figure*}
\centering
\includegraphics
[width=0.95\textwidth]{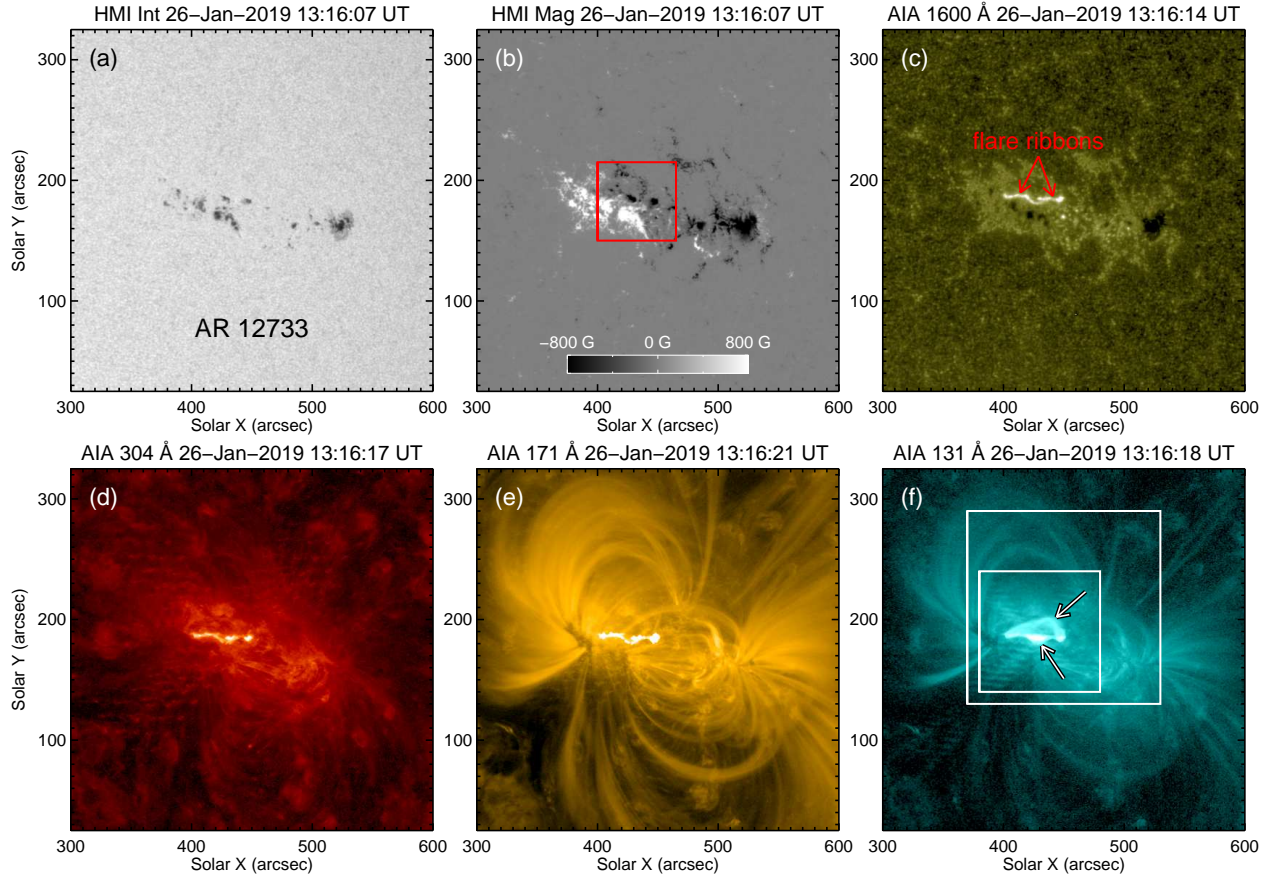} \caption{Overview of AR 12733 at $\sim$ 13:16 UT, 26 January 2019. \emph{Panels (a)-(b)}: HMI intensitygram and LOS magnetogram, respectively. The square in panel (b) outlines the FOV of Figures 2(a)-(c) and Figures 2(e)-(f).  \emph{Panels (c)-(f)}: AIA 1600 {\AA}, 304 {\AA}, 171 {\AA}, and 131 {\AA} images, respectively. In panel (f), the small and large squares outline the FOVs of Figures 4 and 5, and the upper and lower arrows point to the flux rope and post-flare loops, respectively. \protect\\\emph{An animation of this figure is available.}
\label{fig}}
\end{figure*}

\begin{figure*}
\centering
\includegraphics
[width=0.95\textwidth]{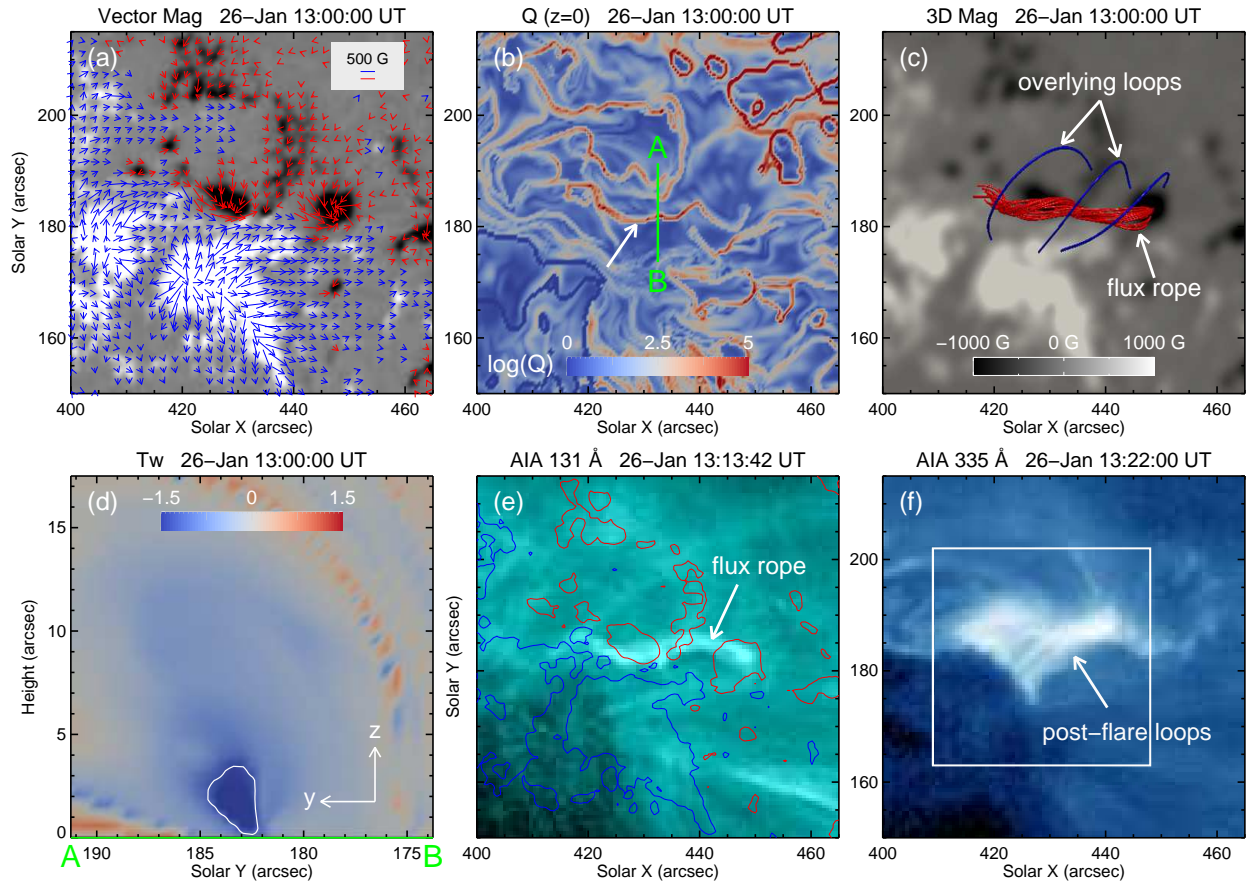} \caption{ \emph{Panel (a)}: HMI photospheric vector magnetogram at 13:00 UT. \emph{Panel (b)}: $Q$ map in the photospheric layer deduced from the NLFFF extrapolation. \emph{Panel (c)}: top view of the NLFFF extrapolated magnetic flux rope (red curves; $|\mathcal{T}_w| > 1$) and the small-scale overlying loops (blue curves). \emph{Panel (d)}: vertical cut of the $\mathcal{T}_w$ distribution along ``A--B" marked in panel (b). The white curve is the contour of the $\mathcal{T}_w$ at $-$1.2 level. \emph{Panels (e) and (f)}: AIA 131 {\AA} image at 13:14 UT when the flare just started soon and AIA 335 {\AA} image at 13:22 UT when the flare peaked, respectively. The blue and red curves in panel (e) are the contours of the LOS magnetic fields at 200 G and $-$200 G, respectively. The square in panel (f) outlines the FOV of Figures 3(a)-(c).
\label{fig}}
\end{figure*}

\begin{figure*}
\centering
{\subfigure{\includegraphics[width=0.95\textwidth]{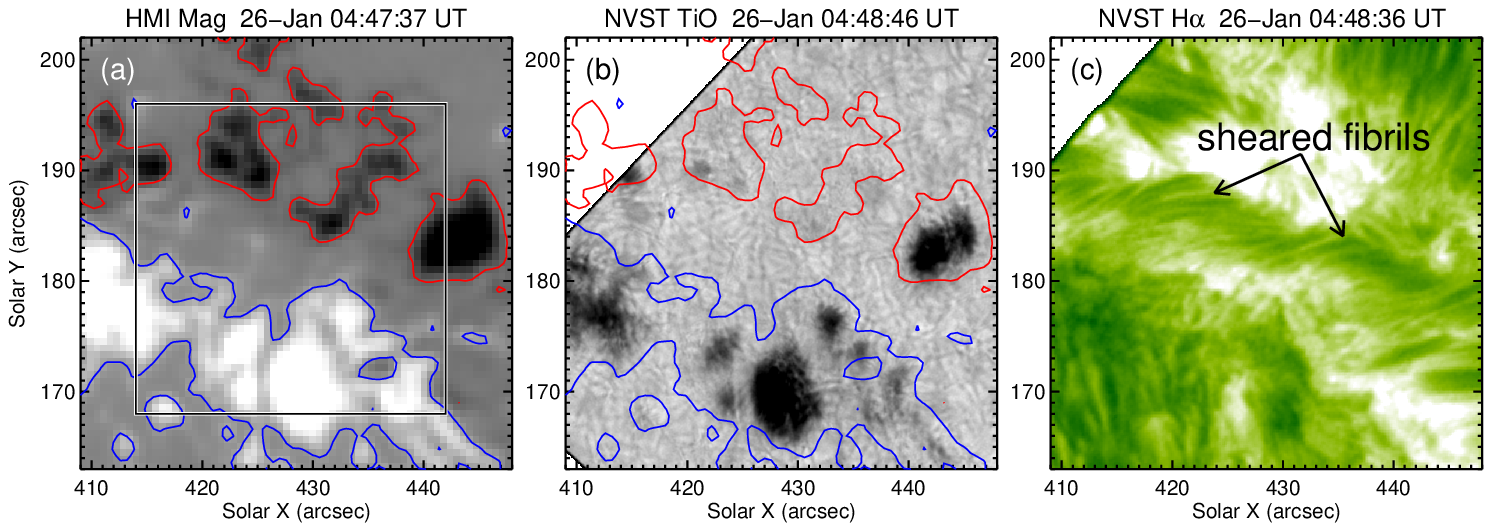}}
\quad
\subfigure{\includegraphics[width=0.95\textwidth]{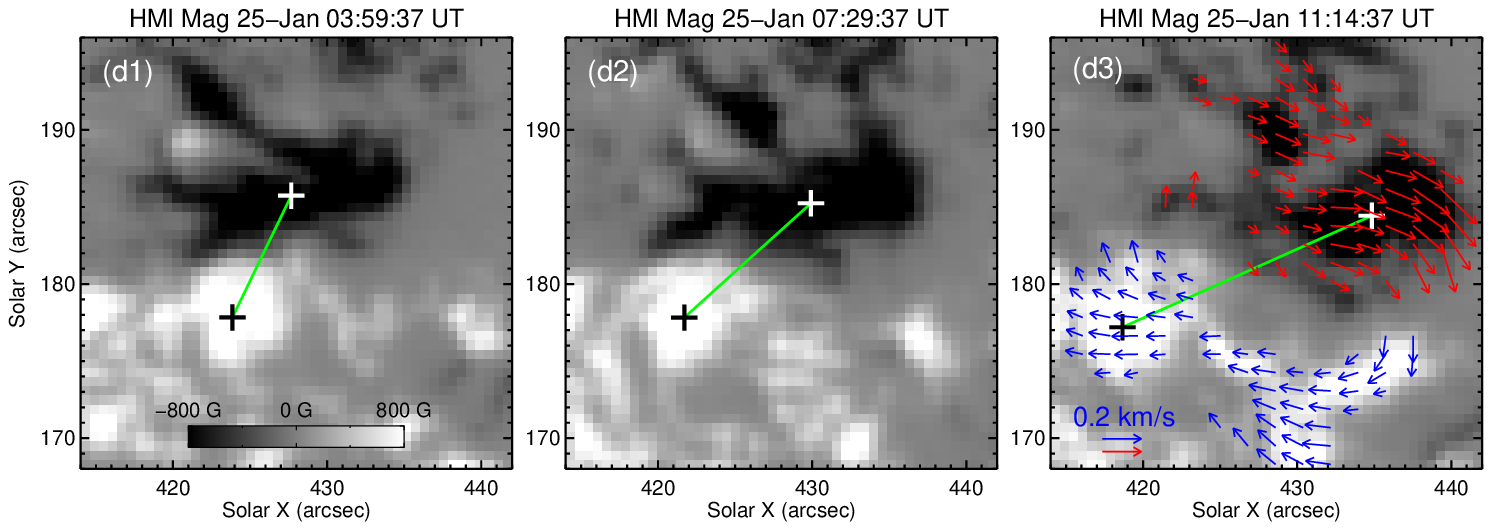}}}
\caption{ \emph{Panels (a)-(c)}: HMI magnetogram, NVST TiO image, and NVST H$\alpha$ image showing the phtospheric magnetic field and intensity map, and the chromospheric fibrils, respectively. The blue and red curves are the contours of the LOS magnetic fields at 200 G and $-$200 G, respectively. The square in panel (a) outlines the FOV of panels (d1)-(d3). \emph{Panels (d1)-(d3)}: sequence of HMI magnetograms showing the shearing motions of the photospheric magnetic fields. The black and white ``+" symbols mark the magnetic centroids of the positive and negative polarities, respectively. The blue and red arrows in panel (d3) indicate the horizontal velocities of the positive and negative magnetic fields, respectively.  \label{fig}}
\end{figure*}

\begin{figure*}
\centering
{\subfigure{\includegraphics[width=0.95\textwidth]{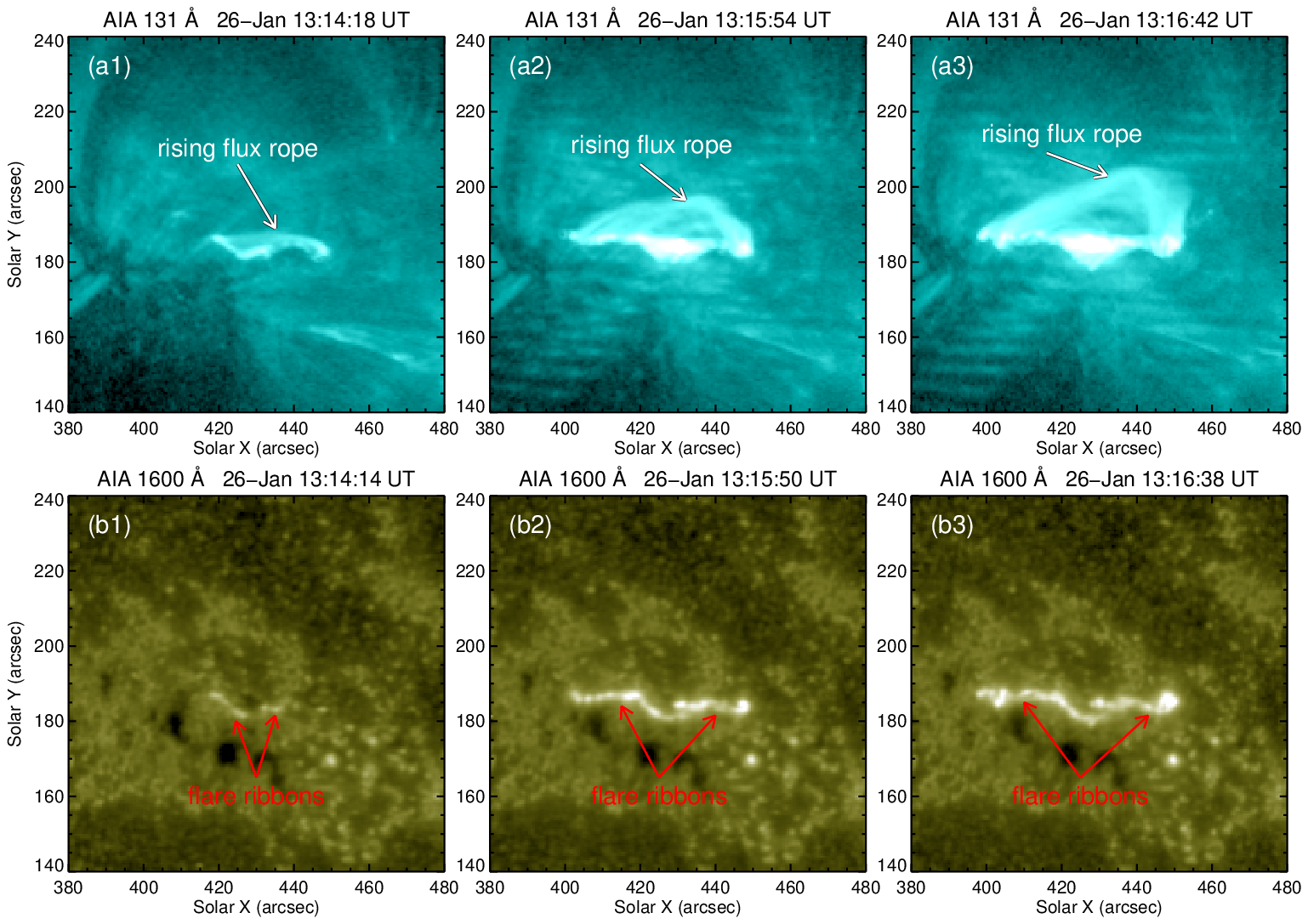}}
\quad
\subfigure{\includegraphics[width=0.95\textwidth]{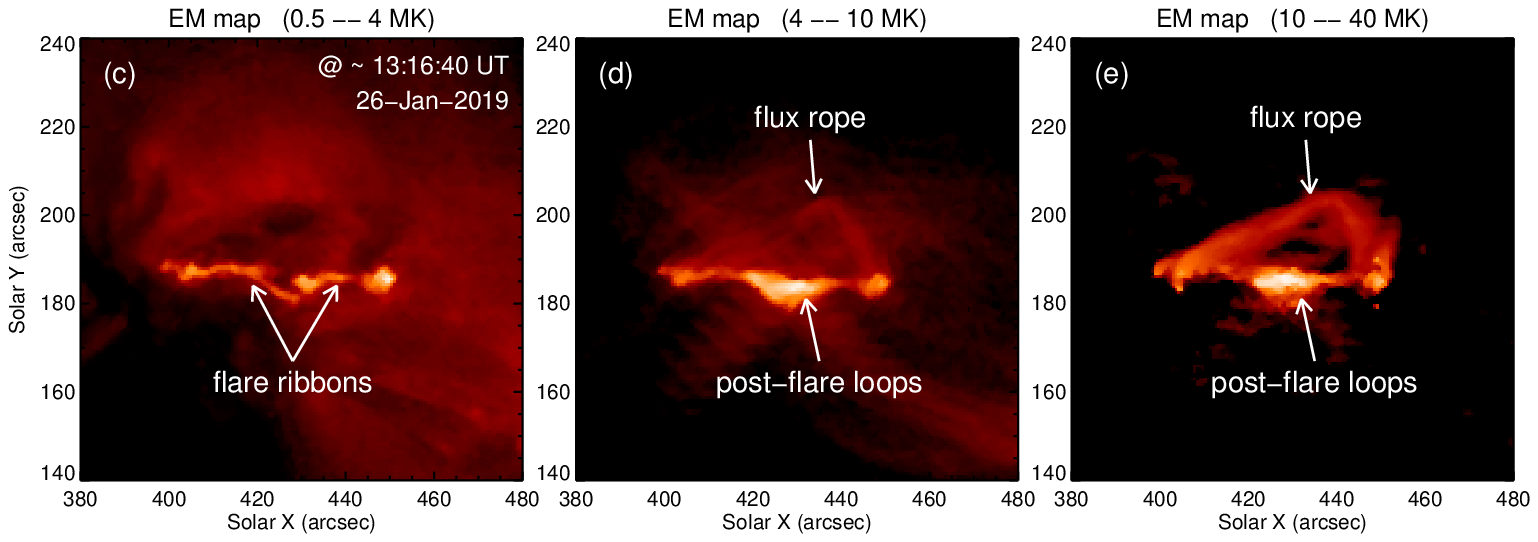}}}
\caption{ \emph{Panels (a1)-(a3)}: AIA 131 {\AA} images showing the rising process of the flux rope at the early stage. \emph{Panels (b1)-(b3)}: corresponding AIA 1600 {\AA} images showing the evolution of two flare ribbons. \emph{Panels (c)-(e)}: EM maps at different temperature ranges at $\sim$ 13:16:40 UT on 26 January. \label{fig}}
\end{figure*}

\begin{figure*}
\centering
\includegraphics
[width=0.95\textwidth]{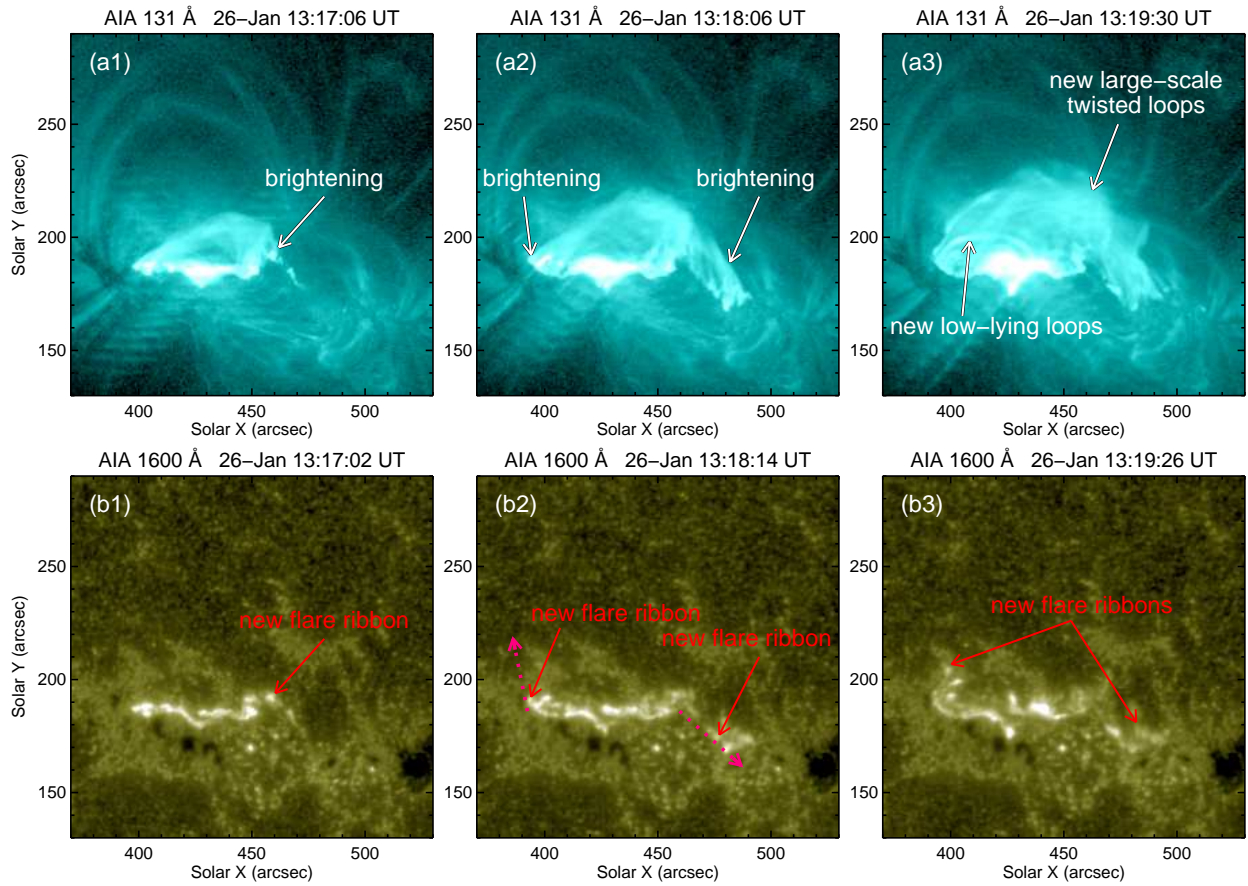} \caption{ \emph{Panels (a1)-(a3)}: AIA 131 {\AA} images showing the reconnection process between the rising flux rope and the large-scale overlying loops. \emph{Panels (b1)-(b3)}: AIA 1600 {\AA} images displaying the evolution of the corresponding flare ribbons. The dotted arrows in panel (b2) indicate the propagation directions of the newly formed flare ribbons.
\label{fig}}
\end{figure*}

\begin{figure*}
\centering
\includegraphics
[width=0.7\textwidth]{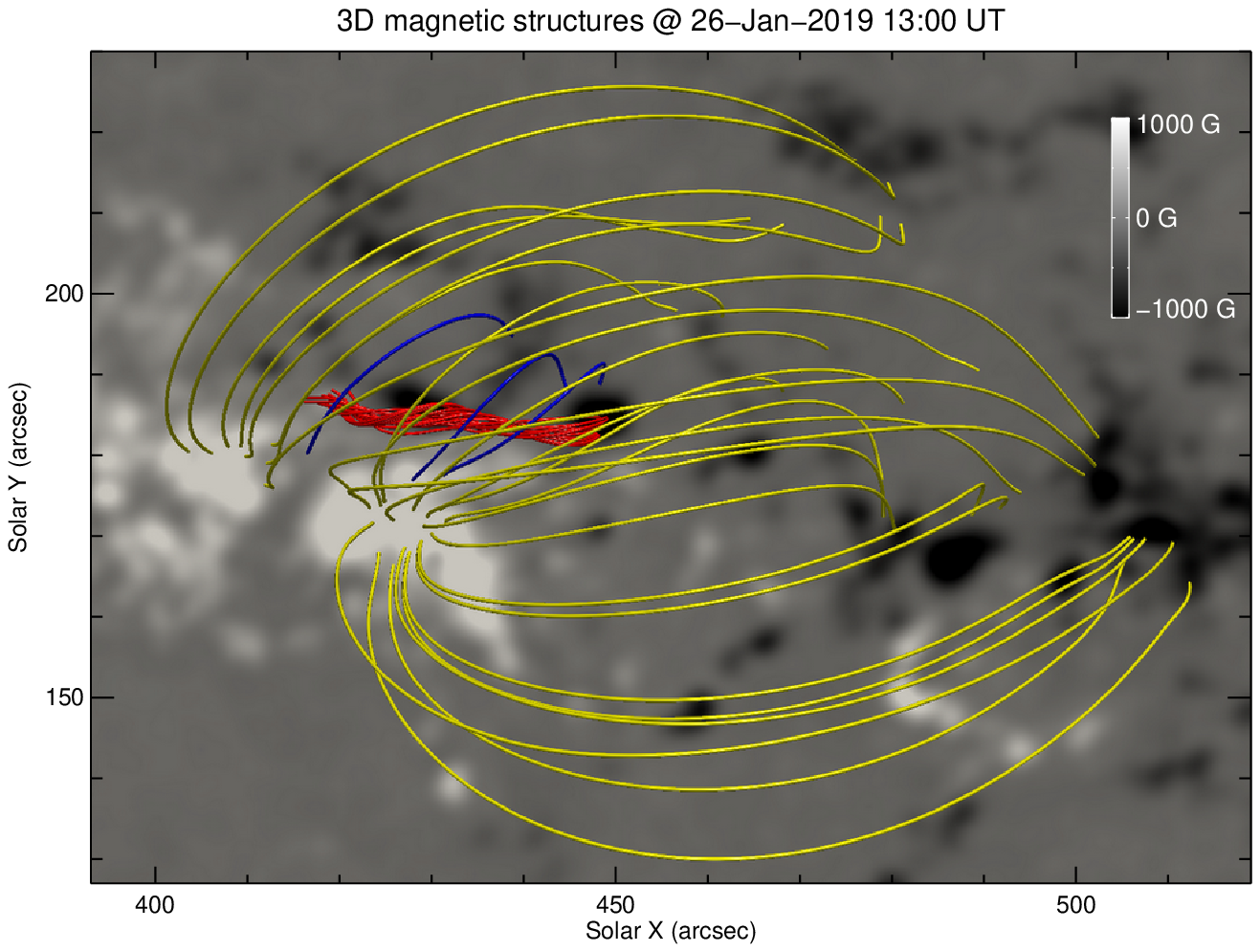} \caption{ Top view of the NLFFF extrapolated 3D magnetic flux rope (red), small-scale (blue) and large-scale (golden) overlying loops in a large FOV.
\label{fig}}
\end{figure*}

The Atmospheric Imaging Assembly (AIA; Lemen et al. 2012) on board the \emph{Solar Dynamics Observatory} (\emph{SDO}; Pesnell et al. 2012) provides the multi-wavelength images of the Sun with the spatial sampling of 0.{\arcsec}6 pixel$^{-1}$. The Helioseismic and Magnetic Imager (HMI; Scherrer et al. 2012; Schou et al. 2012) on board the \emph{SDO} measures the full-disk line-of-sight (LOS) magnetic field and intensity of the photosphere in 6173 {\AA} with a pixel size of 0.{\arcsec}5. Since the C5.0-class flare in active region (AR) 12733 studied here occurred at 13:12 UT, peaked at 13:22 UT, and ended at 13:34 UT on 26 January 2019, the simultaneous AIA and HMI observations from 13:00 UT to 14:00 UT are adopted. We use the extreme-ultraviolet (EUV) images in 304, 171, 193, 211, 335, 94, and 131 {\AA} with a cadence of 12 s, the UV images in 1600 {\AA} with a cadence of 24 s, and the HMI magnetograms and intensitygrams with a cadence of 45 s. In order to study the long duration evolution of the photospheric magnetic field, we employ the LOS magnetograms from 25 January 00:00 UT to 26 January 15:00 UT with a cadence of 3 min. All the \emph{SDO} data are first calibrated from Level 1 to Level 1.5 and then de-rotated differentially to the reference time of 26 January 13:30 UT.

HMI also provides the inverted and disambiguated full-disk vector magnetograms with the spatial sampling of 0.{\arcsec}5 pixel$^{-1}$. We transform the vector magnetic field at each pixel in the image plane to the heliographic components with the formulae given by Gary \& Hagyard (1990), and perform the geometric mapping into the heliographic coordinates. With the photospheric vector magnetogram as the bottom boundary where AR 12733 is located at the center, we extrapolate the coronal magnetic structures using the nonlinear force-free field (NLFFF) modeling (Wheatland et al. 2000; Wiegelmann 2004), before which the photospheric magnetogram is preprocessed to best suit the force-free conditions (Wiegelmann et al. 2006). The calculation is performed within a cubic box of 512 $\times$ 256 $\times$ 128 uniform grid points with $\Delta x= \Delta y=\Delta z=0.$\arcsec$5$. Using the extrapolated three-dimensional (3D) magnetic fields, we also calculate the squashing factor $Q$ (D{\'e}moulin et al. 1996; Titov et al. 2002) and the twist number $\mathcal{T}_w$ (Berger \& Prior 2006) using the code developed by Liu et al. (2016b).

The New Vacuum Solar Telescope (NVST; Liu et al. 2014) is the primary facility of the \emph{Fuxian Solar Observatory} in Yunnan Province of China. The NVST was pointed to AR 12733 from 03:58 UT to 09:21 UT on 26 January 2019 to image the photospheric and chromospheric structures at high tempo-spatial resolution. The TiO 7058 {\AA} images have a pixel size of 0.{\arcsec}052 and a cadence of 30 s. The NVST H$\alpha$ 6562.8 {\AA} images have a pixel size of 0.{\arcsec}136 and a cadence of 9 s. Their fields of view (FOVs) are 124{\arcsec} $\times$ 100{\arcsec} and 126{\arcsec} $\times$ 126{\arcsec}, respectively. They are first calibrated from Level 0 to Level 1 with flat field correction and dark current subtraction, and further reconstructed to Level 1+ by speckle masking (Weigelt 1977; Lohmann et al. 1983). Then we co-align the NVST and \emph{SDO} images by cross-correlating specific features.

\section{Results}

The overview of AR 12733 is shown in Figure 1. The images were taken at $\sim$ 13:16 UT, about 4 min after the start of the C5.0 confined flare. The leading and trailing sunspots (panel (a)) have negative and positive polarities (panel (b)), respectively. The initiation site of the flare is outlined by the red square. Since the flare had started, two flare ribbons were evident in AIA 1600 {\AA} (panel (c)), 304 {\AA} (panel (d)), and 171 {\AA} (panel (e)) images. In the 171 {\AA} (panel (e)) and 131 {\AA} (panel (f)) images, there were lots of large-scale loops connecting the AR's leading and trailing sunspots. Particularly, in the 131 {\AA} image, there were a set of low-lying bright post-flare loops (denoted by the lower arrow), above which a flux rope (appearing as a bright arch-shaped structure as indicated by the upper arrow) can be identified.

The vector magnetic field at the initiation site of the flare is presented in Figure 2(a). The magnetogram was taken at 13:00 UT, i.e., at the pre-flare stage. The magnetic fields near the main polarity inversion line (PIL) are highly sheared. Panel (b) displays the $Q$ map in the photospheric layer (i.e., $x-y$ plane with $z = 0$) deduced from the NLFFF extrapolation. We find that the main PIL of the photospheric magnetic fields well coincides with the high $Q$ region, as denoted by the white arrow, implying that there exists a quasi separatrix layer. We examine the NLFFF extrapolated magnetic fields, and find that, along the PIL, there exists a twisted flux rope (indicated by the red lines in panel (c)) with $\mathcal{T}_w > 1$. Above the flux rope, there are many overlying loops as represented by the blue curves. We make a vertical cut of the $\mathcal{T}_w$ data at the position ``A--B" (as marked in panel (b)), and present the $y-z$ plane $\mathcal{T}_w$ map in panel (d). The high $\mathcal{T}_w$ (dark blue) region coincides with the flux rope above the PIL. The white curve is the contour of the $\mathcal{T}_w$ at $-$1.2 level, and the maximum twist within the flux rope is as high as $-$1.76. At 13:13:42 UT, a brightening structure in 131 {\AA} (denoted by the arrow in panel (e)) was identified to be consistent with the extrapolated flux rope, which could not be detected in any AIA images before the flare. About 8 min later, the flare caused by the flux rope eruption reached its peak, and many post-flare loops (indicated by the arrow in panel (f)) appeared (see also the associated animation).

At the location of the flux rope, we study the lower atmospheric evolution at the earlier time, as shown in Figure 3. Panels (a)-(c) show the appearance of the HMI magnetogram, NVST TiO intensity map, and NVST H$\alpha$ image at around 04:48 UT on 26 January, respectively. We can see that the positive magnetic fields were at the south-east to the main PIL, and the negative at the north-west (panel (a)). The magnetic patches with high field strength coincided with several sunspots observed in the TiO image (panel (b)). In the H$\alpha$ image, there were lots of highly sheared chromospheric fibrils (dark structures denoted by the arrows in panel (c)) connecting the opposite-polarity fields in the two sides of the PIL. In the sequence of photospheric magnetograms (lower panels), significant shearing motions are presented. The magnetic centroids of the positive and negative polarities are marked by the black and white ``+" symbols, respectively. We find that the positive polarity moved towards the left while the negative polarity towards the right. The distance between the centroids of opposite polarities increased, and the orientation from the positive magnetic centroid to the negative one increased clockwise. Using the differential affine velocity estimator (DAVE; Schuck 2006) method, we calculate the photospheric horizontal velocities of the magnetic fields. The horizontal velocities are overlaid in panel (d3), where the blue arrows mainly point to the higher-left and the red arrows to the lower-right. The velocity map definitely shows the existence of shearing motions, which are deemed to play an important role in creating the sheared arcades and the eventual flux rope.

The flux rope eruption only can be observed in the AIA high-temperature passbands, especially in 131 {\AA} channel. Figures 4(a1)-(a3) show the rising process of the flux rope during the early stage of the C5.0 flare. At 13:14:18 UT, the flux rope slightly rose up (panel (a1)), and there was a pair of faint flare ribbons (indicated by the arrows in panel (b1)) in 1600 {\AA} image. About 1.5 min later, the height of the flux rope had significantly increased, as shown in panel (a2), and the flare ribbons in the lower atmosphere elongated accordingly (see panel (b2)). The flux rope continued rising, and 48 s later, it reached to a new height (panel (a3)). We note that, compared to the left section of the rope body, the right section was much higher. In the corresponding 1600 {\AA} image, the bright flare ribbons became more conspicuous (panel (b3)). Moreover, we investigate the temperature properties of the flux rope with the differential emission measure (DEM) method by employing the sparse inversion code (Cheung et al. 2015; Su et al. 2018). A set of AIA images taken at 13:16:40 UT in the 6 channels (i.e, 94, 131, 171, 193, 211, and 335 {\AA}) are used, and the EM maps at different temperature ranges are shown in panels (c)-(e). In the EM map at the temperature range of 0.5--4 MK (panel (c)), the bright structures mainly coincide with the flare ribbons. While in the EM map at the range of 4--10 MK (panel (d)), the bright structures are mainly the post-flare loops, besides which a faint bright structure can be found to coincide with the flux rope. When the temperature is as high as 10--40 MK (see the EM map in panel (e)), the upper structure corresponding to the flux rope is much brighter than that at the lower temperature.

At 13:17:06 UT, about 5 min later after the flare initiation, some brightening structures (denoted by the arrow in Figure 5(a1)) in 131 {\AA} appeared to the west of the rope's right leg, which was also identifiable in the 1600 {\AA} image (panel (b1)). One minute later, more brightening loops successively appeared to the south-west direction (denoted by the right arrow in panel (a2)), and the newly formed flare ribbon propagated to the south-west (directed by the right dotted arrow). Meanwhile, new brightening loops and the corresponding flare ribbon were also observed at the east side of the rope's left leg (see panels (a2) and (b2)). The left newly formed ribbon propagated towards the north-east direction (directed by the left dotted arrow). At 13:19:30 UT, a set of large-scale twisted loops had been formed (see panel (a3)). Moreover, several newly formed low-lying loops were clearly observed. The landscape of the 3D coronal structures in a large FOV is presented in Figure 6. Above the system consisting of the flux rope (the red structure) and the small-scale overlying loops (the blue curves), there are a great number of large-scale field lines (the golden curves) connecting the leading and trailing sunspots, consistent with the observed large-scale loops in 171 {\AA} image (see Figure 1(e)). The external reconnection between the rising flux rope and a part of the large-scale overlying loops is considered to result in the observed newly formed loops and propagating ribbons.

\section{Conclusions and Discussion}

\begin{figure*}
\centering
\includegraphics
[width=0.85\textwidth]{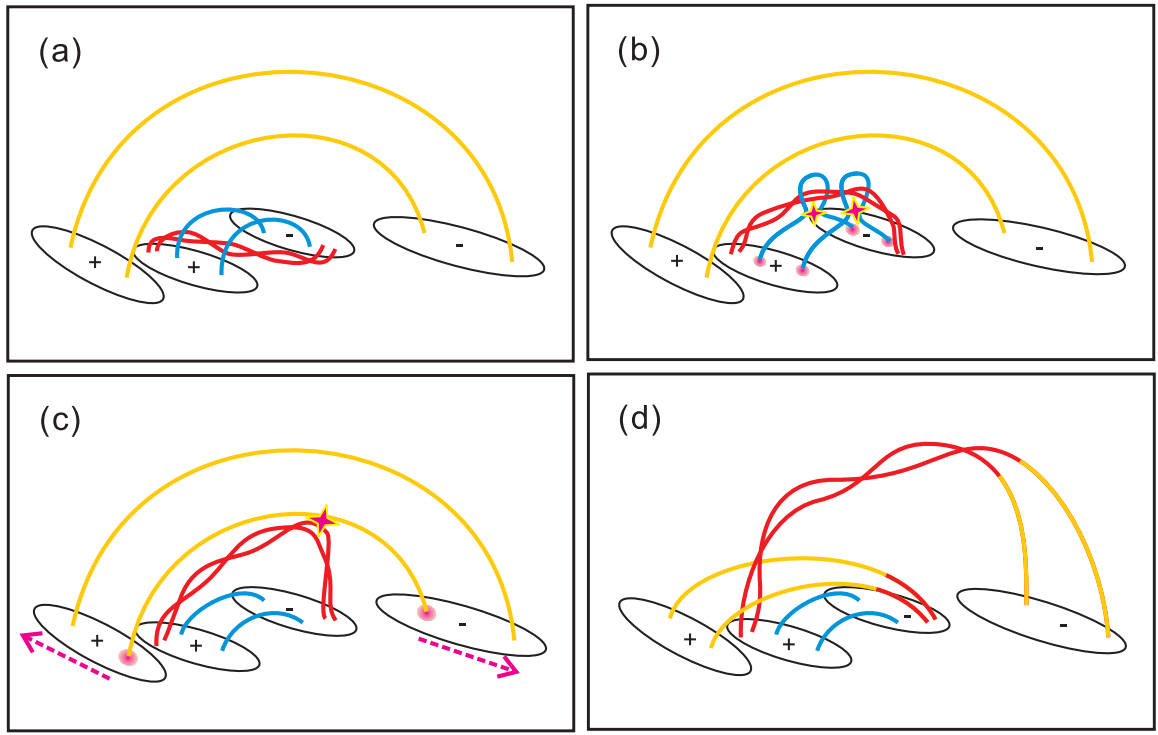} \caption{ Schematic drawings illustrating the two-step (\emph{panels (a)-(b)}: standard rising; \emph{panels (c)-(d)}: external reconnecting) evolution of a magnetic flux rope resulting in a confined flare. The red, blue, and golden curves represent the flux rope, small-scale and large-scale overlying loops, respectively. The star symbols mark the reconnection sites, and the pink dots represent the footpoint brightenings of the reconnecting loops. The two dashed arrows in panel (c) denote the propagation directions of two newly formed flare ribbons.
\label{fig}}
\end{figure*}

Combining the AIA, HMI, and NVST observations, we studied a C5.0 confined flare which occurred in AR 12733 on 26 January 2019. The flare was triggered by a rising magnetic flux rope within the trailing sunspots. At the pre-flare stage, the flux rope was lying above the sheared PIL, but could not be detected in the AIA images. We find that at the earlier time, there were conspicuous photospheric shearing motions between the opposite-polarity fields at the to-be-flaring region, which may be responsible for the flux rope formation. The maximum twist of the flux rope was as high as $-$1.76, and then the flux rope rose and the flare was initiated. Only when the flare started can the flux rope be observed in the high-temperature passbands. The results from the DEM diagnosis confirmed that this flux rope was a high-temperature structure. As the flux rope rose up, there appeared many post-flare loops and two flare ribbons. When the rising flux rope met and reconnected with the large-scale overlying loops, a group of large-scale twisted loops were formed, and a pair of flare ribbons propagating in opposite directions appeared on the outskirts of the former ribbons.

The shearing motion and the flux cancellation in the photosphere have been found to be important in flux rope formation, since the shearing motion can twist the field lines and the flux cancellation associated with tether-cutting reconnection can make the sheared arcades into longer twisted rope (e.g., Moore et al. 2001; Yan et al. 2015). In the present study, before the flare, the photospheric magnetic fields were indeed observed to perform conspicuous shearing motions, and many greatly sheared arcades appeared above PIL, which eventually resulted in the flux rope formation. It should be noted that the flux rope at the pre-flare stage could not be detected in the AIA multi-wavelength images and was traced out when the flare started, which is consistent with the observations of Li et al. (2013) and Yang et al. (2014b). The DEM results reveal that this flux rope was a high temperature structure (see Figure 4(e)), which is in agreement with the results of Zhang et al. (2012) and Cheng et al. (2012) that magnetic flux ropes appeared as hot channels in observations. As shown in Figures 5(b1)-(b3), the propagating directions of two new ribbons were oppositely directed, similar to the observations of Li \& Zhang (2014), implying that the rising flux rope was reconnecting with the large-scale overlying loops, which can also be proved by the formation of the large-scale twisted loops and the new low-lying loops (see Figure 5(a3)).

As revealed by the NLFFF calculation results, the maximum twist number of the flux rope at 13:00 UT on 26 January was as high as $-$1.76. Generally, the threshold value of kink instability for a twisted flux rope is $|\mathcal{T}_w| = 1.75$ (T{\"o}r{\"o}k \& Kliem 2003). Considering the observed twist number exceeded the threshold, the flux rope eruption seems to be initiated by the kink instability. According to the previous studies, when a flux rope rises due to the kink instability, the flux rope converts its twist into writhe or buckles which are typical tell-tale signs that the kink instability results in the destabilization of the flux rope (e.g. Figures 3(e1)-(e3) in Yang et al. 2017). However, in the present study, we did not identify this kind of kinking behavior of the rising flux rope. Hence there may be some other initiating mechanism, such as some adjacent reconnection. As shown in the animation, there are obvious brightenings (outlined by ellipses) in the 304 {\AA}, 171 {\AA}, and 131 {\AA} images, starting at about 13:04:45 UT, peaking at about 13:09:30 UT, and then fading until the start of the main flare. The brightenings occurred in the loops connecting the negative-polarity leading spot on the west with the positive-polarity following spots on the east. The fact that the loops in this vicinity became activated and began to emit first, followed almost immediately by the major flare to the east, implies that there may be some reconnection occurring between adjacent, non-parallel coronal loops at that earlier time. The close correlation in time and the apparent connection of the activated loops to the neighborhood of the flux rope suggest that this reconnection would alter the fields adjacent to or overlying the PIL where the flux rope was formed and could contribute to the flux rope's destabilization.

During the analyzed C5.0-class flare, the altitude of the emitting loops (observed in 131 {\AA}) associated with the flux rope increased with time, and the longitudinal (solar-X) extent of the emitting loops also increased with time, as did that of the flare ribbons (observed in 1600 {\AA}). We tend to interpret this behavior as the consequence of the flux rope's rise. However, besides the possible explanation that the flux rope itself was rising, an alternative explanation for this behavior is that the reconnecting magnetic flux was changing, from loops lying lower in the atmosphere to loops lying higher. The animation shows loops in 171 {\AA} and 131 {\AA} that extend from very near the flux rope and the initiation point of the flare over to the main negative-polarity leading sunspot on the west side of AR 12733. These loops did not move noticeably at the beginning of the flare, and seemed fixed until about 13:20 UT, well after the altitude of the flare-brightened loops had increased dramatically. The current observations are somewhat consistent with the results of Moore et al. (2001), i.e., the reconnection starts in the central core of the flux rope and subsequently propagates out toward the ends of the structure, in this event progressively involving loops that reach to higher heights.

To well describe the eruption process of the flux rope, we give a series of schematic drawings in Figure 7. The flux rope eruption includes two steps: standard rising (panels (a)-(b)) and external reconnecting (panels (c)-(d)). At the pre-flare stage (panel (a)), a low-lying twisted flux rope (red) is overlaid by many small-scale loops (blue), above which there exist lots of much larger field lines (golden). Then the flux rope rises due to the kink instability, and the small-scale overlying loops are stretched upward (panel (b)). Consequently, magnetic reconnection takes place at the sites marked by the star symbols between the oppositely directed loops, and the footpoints of the reconnecting loops are brightened (marked by the pink dots). When the rising rope meets the large-scale overlying field lines, the external reconnection (e.g., Moore et al. 2001; Sterling \& Moore 2001, 2004a, 2004b; Gary \& Moore 2004) begins to take place successively between them (see panel (c)). The two newly formed ribbons propagate in opposite directions as denoted by the pink arrows. Because of the external reconnection, the twist of the flux rope is transferred and spread to the large-scale newly formed system (panel (d)). Our observational results are consistent with the modeling study of DeVore \& Antiochos (2008), in which a confined flare features a flux rope reconnecting with its surrounding environment. They found that the moderately sheared field lines rise rapidly and eventually come to rest in the outer corona instead of escaping the Sun, thus forming the confined eruption.

According to the popular concept, the failed eruption of a magnetic flux rope (appearing as a filament if filled with dark material) contributes to the formation of a confined flare (Ji et al. 2003; T{\"o}r{\"o}k \& Kliem 2005). When a rising flux rope has insufficient energy to break through the confining cage consisting of the overlying loops, it will decelerates and stops at a certain height (Wang \& Zhang 2007; Amari et al. 2018). In the present study, there were a great number of large-scale loops overlying the flux rope as revealed in the AIA 171 {\AA} image (Figure 1(e)) and the extrapolated coronal fields (Figure 6), which should be crucial for the eventual formation of the confined flare. However, different from the previous studies about confined flares, we observe the occurrence of external reconnection when the rising flux rope met the large-scale overlying loops. Due to the external reconnection, the twist of the flux rope was spread to a much larger newly formed system, which plays an important role in the confined flare formation.

\acknowledgments {We are grateful to the referee for the constructive comments and valuable suggestions. The data are used courtesy of AIA, HMI, and NVST science teams. \emph{SDO} is a mission for NASA's Living With a Star (LWS) Program. This work is supported by the National Natural Science Foundations of China (11673035, 11790304, 41404136, 11773039, 11533008, 41774195, 11790300), Key Programs of the Chinese Academy of Sciences (QYZDJ-SSW-SLH050), Youth Innovation Promotion Association of CAS (2014043 and 2017078), and Young Elite Scientists Sponsors hip Program by CAST (2018QNRC001).\\ }

{}

\clearpage

\end{document}